\documentclass[a4paper,11pt]{article}
\usepackage[english]{babel}
\usepackage[latin1]{inputenc}
\usepackage[T1]{fontenc} 
\usepackage{ae,aecompl} 
\usepackage{amssymb}
\usepackage{amsmath}
\usepackage{graphicx}
\usepackage{subfig}
\usepackage{fancyhdr}
\IfFileExists{.pdfmode}{%
  
  \usepackage{ae,aecompl} 
  }{
  
  }

\begin{document}

\thispagestyle{plain}
\begin{center}
\Huge{Modeling resonant trojan motions}\\
\huge{in planetary systems}
\end{center}

\begin{center}
\Large{Christos Efthymiopoulos}\\
\small{Research Center for Astronomy and Applied Mathematics, Academy of Athens, Greece}\\
\small{\textsc{cefthim@academyofathens.gr}}\\
\vspace{0.3cm}
\Large{Roc\'io Isabel P\'aez}\\
\small{Dip. di Matematica, Universit\`a di Roma ``Tor Vergata''}\\
\small{\textsc{paez@mat.uniroma2.it}}\\
\end{center}

\noindent
{\bf Abstract:} We consider the dynamics of a small trojan companion
of a hypothetical giant exoplanet under the secular perturbations of
additional planets.  By a suitable choice of action-angle variables,
the problem is amenable to the study of the slow modulation, induced
by secular perturbations, to the dynamics of an otherwise called
`basic' Hamiltonian model of two degrees of freedom (planar case). We
present this Hamiltonian decomposition, which implies that the slow
chaotic diffusion at resonances is best described by the paradigm of
modulational diffusion.

\begin{center}
\line(1,0){250}
\end{center}

\section{Introduction}

Despite extensive search, no pairs of co-orbital exoplanets have 
been discovered so far. Some reasons for the unlikeliness of the 
co-orbital configuration are discussed in Giuppone et al. (2012), 
Haghighipour (2013), Dobrovolskis (2013) and Pierens and Raymond 
(2014). Dynamical obstructions appear in the formation process 
as well as during the migration and/or capture of the planets 
into resonance. Besides these constraints, however, there is also 
the question  of the {\it long-term stability } of co-orbital motions. 
This means the stability of the orbits over timescales comparable 
to the age of the hosting system. 

In a recent work (P\'{a}ez and Efthymiopoulos 2014) we initiated 
a study of the long-term stability in a hypothetical configuration 
in which a small (considered massless) planet moves around the 
Lagrangian points of a giant primary. Numerical simulations have 
shown that up to Earth-sized trojan planets can appear close to 
gaseous giants (Beaug\'{e} et al. 2007, Lyra et al. 2009). 
This dynamical system is a case of the elliptic restricted three 
body problem (ERTBP), or, with additional planets, the `restricted 
multi-body problem' (RMBP). Alternative applications of the RMBP 
encompass co-orbital satellites of a planet, asteroids, and 
artificial trojan objects in a Sun-planet or planet-moon system. 

In an accompanying poster (P\'{a}ez and Efthymiopoulos, this volume) 
we outline one of our so-far obtained numerical results, referring to 
the diffusion timescales in the case of initial conditions taken close 
to some so-called {\it secondary resonances} within the co-orbital 
domain. Several authors (e.g. \'{E}rdi et al. 2007, 2009, Schwarz 
et al. 2007) have stressed the importance of secondary resonances 
in the problem of long-term stability. Related numerical works,  
applied to Jupiter's trojan asteroids, are Marzari et al. (2003), 
Robutel and Gabern (2006), Robutel and Bodossian (2009). Our own 
numerical work compares maps of the resonant structure, as depicted 
in a suitably defined domain of action variables (i.e. proper 
elements), with maps of the stability times for initial conditions 
within the resonance web. We found evidence of a tight correlation 
between the two maps (see P\'{a}ez and Efthymiopoulos, this volume).

In the sequel we briefly discuss how our Hamiltonian formulation in 
action-angle variables is introduced in the framework of the RMBP, 
as well as the consequences this formulation leads to regarding the 
dynamical characterization of the problem. 

\section{Summary of the Hamiltonian formulation}

A summary of our formulation is the following: assuming all perturbing 
planets far from mean motion resonances, by a suitable sequence of 
canonical transformations we arrive (in the planar case) at expressing 
the Hamiltonian of the RMBP as:
\begin{equation}\label{hamrmpp}
H=H_b(J_s,\phi_s,Y_f,\phi_f,Y_p;e_0)
+H_{sec}(J_s,\phi_s,Y_f,\phi_f,Y_p,\phi,P_1,\phi_1,...,P_S,\phi_S)~~.
\end{equation}
i) The pairs $(Y_f,\phi_f)$, $(J_s,\phi_s)$, $(Y_p,\phi)$ are  
action-angle conjugate variables corresponding to the `short-period', 
`synodic' and `secular' motions of the trojan body respectively. 
The short-period terms correspond physically to epicyclic oscillations. 
The synodic oscillations describe the `long period' librations around 
the Lagrangian points L4 or L5. The action variable $J_s$ determines 
the value of the `proper libration' (see Milani (1993), or Beaug\'{e} 
and Roig (2001) for the definition of trojan proper elements). 
The action $Y_p$ labels the `proper eccentricity'. The angle 
$\phi$ measures phase oscillations around an angle $\beta$ (see
below) which expresses the relative difference between the arguments 
of perihelia of the trojan body and the giant primary. We note that 
an analysis omitted here allows to see that the form of the Hamiltonian 
(\ref{hamrmpp}) implies that the oscillations of $\beta$ are bounded. 
Finally, the pairs $(\phi_i,P_i)$, $i=1,...,s$ are action-angle 
variables for the oscillations of the eccentricity vectors of 
the $S$ additional planets. 

ii) We call the first term $H_b$ in (\ref{hamrmpp}) the `basic 
model'. The angle $\phi$ is ignorable in $H_b$, implying that 
$Y_p$ is a constant of motion under the dynamics of $H_b$ alone. 
The parameter $e_0$ is the mean modulus of the eccentricity vector 
of the giant primary. Thus, $H_b$ represents a system of two degrees 
of freedom, wherein both $e_0$ and $Y_p$ act as parameters, 
i.e. the `forced' ($e_0$) and `proper' ($e_p=\sqrt{-2Y_p}$) 
eccentricity. 

iii) The term $H_{sec}$ contains only trigonometric terms depending 
on the slowly varying angles $\phi$,$\phi_i$, $i=1,...,s$. Hence, 
$H_{sec}$ introduces only secular perturbations to the dynamics 
under $H_b$. In particular, $H_{sec}$ causes a slow pulsation of 
the chaotic separatrix-like layers at the borders of the resonances 
arising under $H_b$. As shown in P\'{a}ez and Efthymiopoulos 2014, 
this phenomenon is best described by the paradigm of `modulational 
diffusion' (Chirikov et al. 1985). 

iv) The form of the function $H_b$ is identical in the ERTBP and 
the RMBP, setting $e_0=e'$ and $\beta=\omega$ in the former, where 
$e'$ is the (constant) eccentricity of the primary, and $\omega'=0$ 
its pericentric position. This formal equivalence implies that the 
qualitative features of the diffusion along resonances, as they 
appear in the plane of the action variables $J_s,Y_p$, are similar 
in the RMPP and the ERTBP. Examples of the latter are studied in 
P\'{a}ez and Efthymiopoulos (2014). 

We now summarize the derivation of the Hamiltonian (\ref{hamrmpp}).  
We assume that, far from mean-motion resonances, the time evolution 
of the eccentricity vectors of all massive bodies can be approximated 
by quasi-periodic formulae 
\begin{eqnarray}\label{excvec}
e'\exp{i\omega'}&=&e_0'\exp{i(\omega_0'+g't)}+\sum_{k=1}^{S} A_k
\exp{i(\omega_{k0}'+g_kt)}\nonumber\\
e_j\exp{i\omega_j}&=&B_{j0}\exp{i(\omega_{0j}+g't)}+\sum_{k=1}^{S}
B_{kj} \exp{i(\omega_{kj}'+g_kt)}~~
\end{eqnarray}
\noindent
setting, without loss of generality, $\omega_{0}'=0$. The constants 
$g'$, and $g_j$, $j=1,\ldots s$ are secular frequencies associated 
with the primary and the $S$ planets respectively. Also, we assume 
that the condition $e_0' > \sum_{k=1}^{S} A_k$ holds for the giant 
primary, implying an average constant rate of precession of its 
perihelion with frequency $g'$. One has $e' = e_{0}' + F$, 
$\omega' = \phi' + G$, where $\phi'=g't$ and $F$ and 
$G$ are of first order in the amplitudes $A_k$, $k=1,...,s$. 
Averaged over the mean longitudes $\lambda_1,\ldots,\lambda_S$ 
the Hamiltonian reads
\begin{equation}\label{hamsec}
H=-{1\over 2(1+x)^2} + I_3 + g'I'+ \sum_{j=1}^S g_j I_j - 
\mu R(\lambda,\omega,x,y,\lambda',\phi';e_0') 
- \mu R_2 - \sum_{j=1}^S\mu_j{\cal R}_{j}
\end{equation}
where: i) $x=\sqrt{a}-1$, $y=\sqrt{a}\left(\sqrt{1-e^2}-1\right)$ 
are Delaunay action variables, $(a,e)$ being the major semi-axis 
and eccentricity of the trojan body (in units in which $a'=1$ for 
the primary), and $(\lambda,\omega)$ the mean longitude and 
argument of the perihelion. The variables $I_3$, $I'$, $I_j$, 
$j=1,\ldots,S$ are dummy actions congugate to the angles 
$\lambda'$, $\phi'=g't$ and $\phi_j=g_j t$. 
ii) $R$ is has the same form as the disturbing function  
in the ERTBP with the substitution $e_0\rightarrow e'$, 
$\phi'\rightarrow\omega'$, with $\mu$ equal to the primary's 
mass parameter (all functions and variables are considered 
in the heliocentric frame). 
iii) $R_2$, expressing the indirect effects of the $S$ 
additional planets,  comes from replacing $e'=e_0'+F(\phi',\phi_j)$, 
$\omega'=\phi'+G(\phi',\phi_j)$ in the disturbing function 
of the ERTBP and Taylor-expanding around $e_0'$ and $\phi'$, 
assuming $F$ and $G$ small quantities. 
iv) Finally, ${\cal R}_{j}$ are the (averaged over mean 
longitudes) direct terms of the $S$ additional planets. 

The canonical transformation $\tau=\lambda-\lambda'$, 
$\beta=\omega-\phi'$, $J_3=I_3+x$, $P'=I'+y$ allows 
to re-express the hamiltonian in terms of the resonant 
angle $\tau$ and the relative argument of pericenter 
difference $\beta$. The Hamiltonian can be recast as 
$H=<H>+H_1$, where
$$
<H>=-{1\over 2(1+x)^2} -x + J_3 - g'y -\mu <R>(\tau,\beta,x,y;e_0')
~~~~~~~~~~~~~~~
$$
$$
~~~~H_1=
 g'P'+ \sum_{j=1}^S g_j I_j -\mu 
\tilde{R}(\tau,\beta,x,y,\lambda',\phi';e_0')
~~~~~~~~~~~~~~~~~~~
~~~~~~~~~~~~~~~~~~~
$$
$$
-\sum_{j=1}^S\mu_j R_j(x,y,\beta,\phi',\phi_1,...,\phi_s)
- \mu{\cal R}_2(x,y,\tau,\beta,\phi',\phi_1,...,\phi_s)
$$
with 
$<R>={1\over 2\pi}\int_{0}^{2\pi}R d\lambda'$, $\tilde{R}=R-<R>$. 
The Hamiltonian $<H>$ allows to determine the forced equilibrium 
by the solution to the system of equations $\partial<H>/\partial x$ 
$=$ $\partial<H>/\partial y$ $=$ $\partial<H>/\partial \tau$ 
$=$ $\partial<H>/\partial \beta$ $=0$. One finds that
\begin{equation}\label{forced}
\bigg(\tau_0,\beta_0,x_0,y_0\bigg)=
\bigg(\pi/3,\pi/3,0,\sqrt{1-e_0'^2}-1\bigg)+O(g')~~.
\end{equation}
Note that the forced equilibrium represents a relative configuration, 
i.e., the eccentricity vector of the trojan body has the same 
modulus $e_0$ and a constant {\it relative} angle with respect 
to the {\it mean} eccentricity vector of the primary. This result 
follows also by a careful inspection of the formulae provided in 
Morais (2001). 

Expanding around the forced equilibrium, we introduce new variables 
\begin{equation}\label{poincvar}
v=x-x_0,~~u=\tau-\tau_0,~~Y=-(W^2+V^2)/2,~~\phi=\arctan(V/W) 
\end{equation}
$$
V=\sqrt{-2y}\sin\beta-\sqrt{-2y_0}\sin\beta_0,~~~~
W=\sqrt{-2y}\cos\beta-\sqrt{-2y_0}\cos\beta_0~~.
$$
The variables $(v,u)$ describe the motion in the synodic plane, 
while the action variable $Y$ measures the distance of an orbit 
from the forced equilibrium position in the secular plane 
$(\sqrt{-2y}\cos\beta,\sqrt{-2y}\sin\beta)$. Finally, we 
introduce the canonical transformations $Y_p=Y+J_3$, 
$\phi_f=\lambda'-\phi$, and 
\begin{equation}\label{synodic}
J_s={1\over 2\pi}\int_C (v-v_0)d(u-u_0)
\end{equation}
where the integration is over a closed invariant curve $C$ around 
$(u_0,v_0)$, with conjugate angle $\phi_s$. Substituting these 
transformations yields the form (\ref{hamrmpp}) of the Hamiltonian. 

The study of the basic model allows to identify the most important 
secondary resonances, which are commensurabilities between the 
fast and synodic frequencies $\omega_f=\dot{\phi}_f$, $\omega_s
=\dot{\phi}_s$. The fast frequency is related to the secular frequency 
$g=\dot{\phi}$ by $\omega_f=1-g$, in units in which the mean motion 
of the giant primary is equal to 1. The general form of a resonance 
is
\begin{equation}\label{resgen}
m_f\omega_f+m_s\omega_s+m g + m'g'+m_1g_1+\ldots+m_Sg_S=0
\end{equation}
with $m_f,m_s,m,m',m_j$ (with $j=1,\ldots,S$) integers. The resonances 
of the basic model exist in the complete hierarchy of problems, from 
the planar circular restricted three body problem ($s=0$, $g'=0$, 
$e_0'=0$) up to the complete multi-body problem. For the mass parameters 
of giant exoplanets the most important resonances are of the form 
$\omega_f-n\omega_s=0$, with $n$ in the range $4\leq n\leq 12$ for 
typical mass parameters of the gaseous primary. In the frequency space 
$(\omega_f,\omega_s,g)$, these resonances define planes normal to the 
plane $(\omega_f,\omega_s)$ which intersect each other along the 
$g$--axis. All other resonances with $|m|+|m'|+|m_1|+\ldots+|m_S|>0$ 
intersect transversally one or more planes of the main resonances. 
We refer to such resonances as `transverse' if $|m_f|+|n|>0$, or 
`secular' if $|m_f|+|n|=0$. In P\'{a}ez and Efthymiopoulos (2014), 
we show that the diffusion along transverse or secular resonances 
is of the Arnold type, hence very slow. On the other hand, there 
are transverse resonances which accumulate to multiplets around 
the main ones, thus producing a faster (modulational) diffusion. 

\vspace{0.5cm}
\noindent
\large{{\bf Acknowledgements:}} R.I.P. was supported by the
Astronet-II Training Network (PITN-GA-2011-289240). C.E. was supported
by the Research Committee of the Academy of Athens (Grant 200/815) and
by an IAU Symposium Grant.

\end{document}